\documentclass{PoS}
\usepackage{amsmath}
\usepackage{subfigure} 
\usepackage{wrapfig}
\usepackage{cite}
\allowdisplaybreaks[1]

\newcommand{\del}{\partial}

\newcommand{\q}{{\rm{q}}}

\newcommand{\V}{{\rm V}}
\newcommand{\A}{{\rm A}}

\title{
\begin{picture}(0,0)(0,0)%
   \put(300,75){\makebox(0,0)[l]{\textnormal
{\normalsize KEK-CP-338, OH-HET-879 
}
}}%
\end{picture}%
Analysis of short-distance current correlators using OPE}

\ShortTitle{Analysis of short-distance current correlators using OPE}

\author{\speaker{M. Tomii}$^{a,b}$,
	\ G.~Cossu$^b$, B.~Fahy$^b$, H.~Fukaya$^c$, S.~Hashimoto$^{a,b}$, J.~Noaki$^b$
	(JLQCD~Collaboration)\\
E-mail: \email{tomii@post.kek.jp}\\
	\it
$^a$ Department of Particle and Nuclear Science,
SOKENDAI (The Graduate University for Advanced Studies),
Ibaraki 305-0801, Japan,\\
$^b$ KEK Theory Center, Institute of Particle and Nuclear Studies, High Energy Accelerator Research Organization (KEK), Ibaraki 305-0801, Japan,\\
$^c$ Department of Physics, Osaka University, Toyonaka 560-0043, Japan.\\
        	}

\abstract{
We investigate  the correlators of flavor non-singlet bilinear operators
calculated on the lattice at short distances.
In the continuum theory, non-perturbative effects
are encoded in the form of the operator product expansion (OPE).
We test the prediction of OPE by comparing lattice results with
those in the continuum theory.
We also determine the renormalization factors of quark currents.
}

\FullConference{The 33rd International Symposium on Lattice Field Theory\\
		14 -18 July 2015\\
		Kobe International Conference Center, Kobe, Japan}

\begin{document}

\section{Introduction}

Current correlators provide a
rich source of information on the QCD vacuum.
Correlators in the sufficiently short-distance region ($< 0.1$ fm) are
well predicted by perturbation theory, which is currently available to four-loop
order \cite{Chetyrkin_Maier_2011},
while the non-perturbative dynamics dominates in
the region of long distance ($> 1$ fm), as systematically understood
by chiral perturbation theory.
Dynamics in the middle range is, on the other hand, quite nontrivial,
and lattice calculation is needed to obtain quantitative predictions.

The operator product expansion (OPE) \cite{Shifman:1978bx}
accommodates some non-perturbative effects
and is used to analyze experimental data such as the hadronic $\tau$ decays.
Since the number of operators to be included rapidly increases as one considers
longer distance physics the range of distances suitable for OPE is limited.
In this work, we test the applicability of OPE for
current correlators in the region $0.1$ fm $< x < 0.5$ fm by comparing
current correlators in the continuum theory and lattice calculations.

As a by-product, we can determine the renormalization constants of quark bilinear
operators following the analysis of
\cite{Martinelli_etal:1997,V.Gimenez_etal:2004,K.Cichy_etal:2012},
in which the renormalization condition
is imposed on the correlator at a certain distance in the coordinate space.
Unlike the RI/MOM scheme \cite{Martinelli:1994ty}, this method enables us to
renormalize composite operators in a fully gauge invariant manner
and to use the perturbative matching factor available to the four-loop level.
Like RI/MOM, the window problem remains, {\it i.e.} we need to use the
correlators in the region satisfying
$a\ll x\ll \Lambda_{\rm QCD}^{-1}$ to avoid discretization effects
on the lattice and non-perturbative effects on the continuum side.
We investigate these effects and introduce various techniques to reduce them.

In this report, we present the status of these analyses. We employ $2+1$
flavor M\"{o}bius domain wall fermions with stout link smearing and
the Symanzik improved gauge action.
We work on $32^3\times64$ lattices at $a^{-1} = 2.45$ GeV, $48^3\times96$
lattices at $a^{-1}=3.61$ GeV and a $64^3\times128$ lattice at $a^{-1} = 4.50$ GeV,
all of which have matched physical volume and pion
masses of $M_\pi = 300 \sim 500$ MeV.
Calculation of
masses and decay constants for the light mesons and the heavy-light mesons
on these ensembles is reported in \cite{Fahy_etal:2015}.

\section{Current correlators}

We calculate correlation functions of
light quark 
bilinear operators in the coordinate space,
\begin{equation}
\begin{array}{ll}
\Pi_{\rm S}(x) = \langle S(x)S(0)^\dag\rangle,
& \hspace{10mm}
\Pi_{\rm P}(x) = \langle P(x)P(0)^\dag\rangle,
\\
\Pi_{\V,\mu\nu} (x) = \langle V_\mu(x)V_\nu(0)^\dag\rangle,
& \hspace{10mm}
\Pi_{\A,\mu\nu} (x) = \langle A_\mu(x)A_\nu(0)^\dag\rangle,
\end{array}
\label{eq:def_correl}
\end{equation}
where $S$ and $P$ are the scalar and pseudoscalar densities,
while $V_\mu$ and $A_\mu$ are the vector and axial-vector currents.
The flavor indices are omitted for simplicity, but they are understood as isospin triplet
operators of light quarks.
We also use the vector and axial-vector correlators after taking a trace of
the Lorentz diagonal components,
\begin{equation}
\Pi_{\V/\A}(x) = \sum_\mu\Pi_{\V/\A,\mu\mu}(x).
\label{eq:def_sum_va}
\end{equation}

\begin{figure}[tb]
\begin{center}
\includegraphics[width=9.5cm]{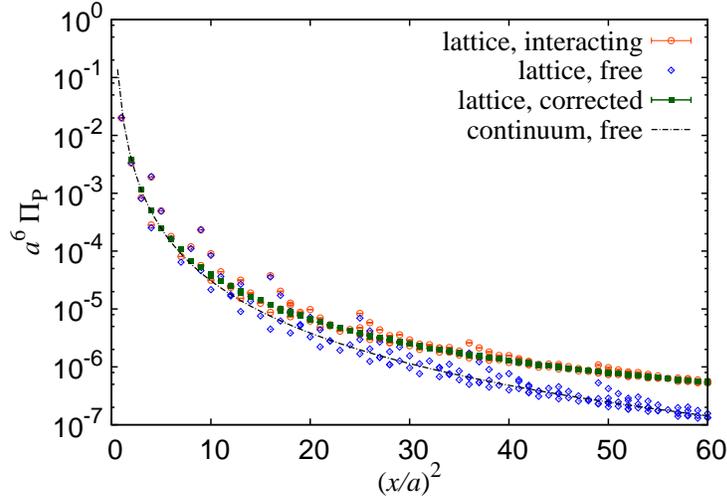}
\caption{
The pseudoscalar correlators measured on the lattice before (circle) and after
(square) subtracting discretization effects in the tree level.
The correlators in the free system calculated
in the continuum theory (dashed curve) and lattice (diamond)
are also plotted.
This data are calculated on the $48^3\times96$ lattice at $\beta=4.35$ and input
mass $am_\q = 0.0042, am_s = 0.0180$.
}
\label{fig:x2vspp}
\vspace{-4mm}
\end{center}
\end{figure}

Figure~\ref{fig:x2vspp} shows the pseudoscalar correlator calculated
non-perturbatively on the lattice (circle) and that in the free system both in the
continuum (dashed curve) and lattice theory (diamond).
Due to the discretization effect, the non-perturbative data are not on a smooth line.
However, the similar effect is already seen in the free correlator, which
implies that the difference of the free theory between the lattice and the continuum
describes the discretization effect in the interacting system to a good approximation.
In fact, by applying a subtraction,
\begin{equation}
\Pi_\Gamma^{\rm lat} \rightarrow
\Pi_\Gamma^{\rm lat} - \big(\Pi_\Gamma^{\rm lat,\ free} - \Pi_\Gamma^{\rm free,\ cont}\big),
\end{equation}
we obtain smoother correlator as shown in Fig.~\ref{fig:x2vspp} by squares.

\begin{figure}[tb]
\begin{center}
\subfigure{\mbox{\raisebox{1mm}{\includegraphics[width=73mm]{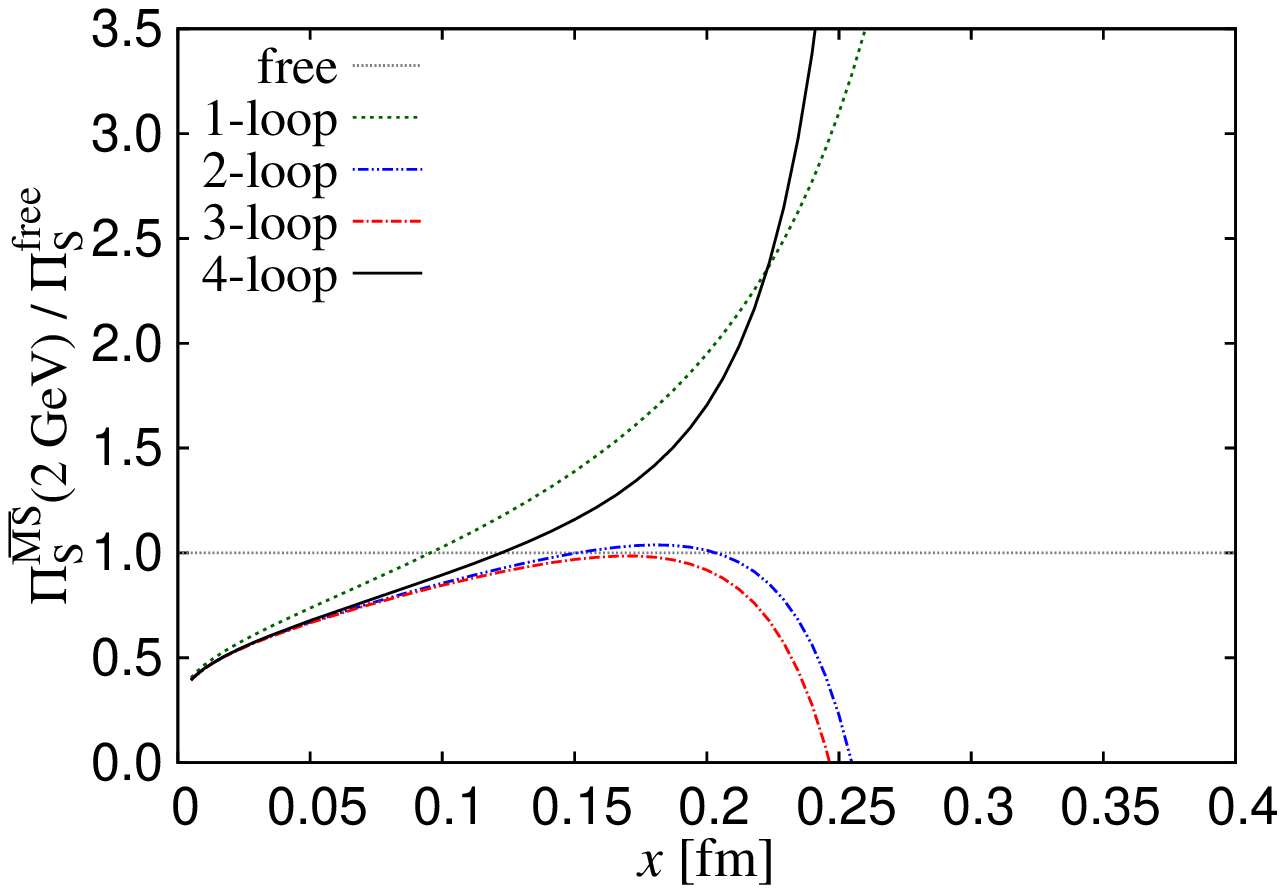}}}}
\subfigure{\mbox{\raisebox{1mm}{\includegraphics[width=73mm]{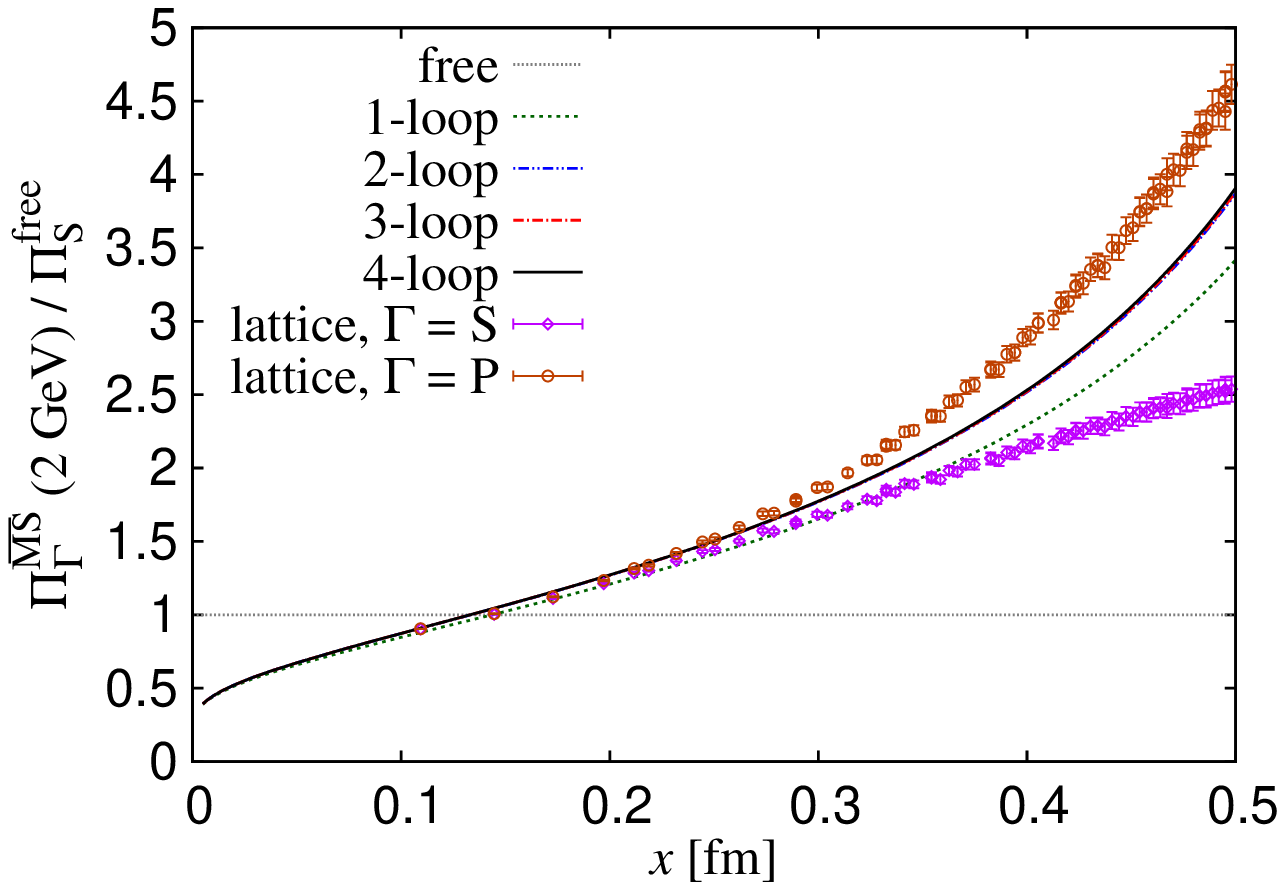}}}}
\caption{
The scalar correlator in the continuum perturbation theory calculated with
the coupling constant $\alpha_s(\mu^*)/\pi$ renormalized at $\mu^* = 1/x$ (left) and $2.86/x$
(right) in the $\rm\overline{MS}$ scheme. 
The correlators are renormalized at 2 GeV in the $\rm\overline{MS}$ scheme
with $n_f = 3$.
In the right panel, we also plot lattice correlators of the scalar and pseudoscalar channles
measured on the same ensemble as Fig.~1.
}
\label{fig:ss}
\vspace{-3.5mm}
\end{center}
\end{figure}

Next, we examine the convergence of the continuum perturbation theory to
identify the valid region of $x$.
It is sufficient to investigate the scalar and vector channels
because these channels coincide with the pseudoscalar and axial-vector
channels, respectively, in the massless limit.
Perturbative coefficients of correlators are calculated to four-loop in
\cite{Chetyrkin_Maier_2011}.
The beta function \cite{Chetyrkin_etal_1997} and the mass anomalous dimension
\cite{K.G.Chetyrkin_1997,J.A.M.Vermaseren_etal:1997}
are also known up to four-loop level.
However, if we na\"\i vely use these results,
the ratio shows poor convergence as shown in the left panel of
Fig.~\ref{fig:ss}.
In this plot, reasonable convergence is found only in the region $x \lesssim 0.1$ fm,
which is in the same order as our lattice spacing $a = 0.04\sim0.08$ fm.
We can improve the convergence of the perturbative expansion by
choosing an appropriate renormalization scale $\mu^*$
instead of using $\tilde\mu \simeq 1.12/x$ suggested in
\cite{Chetyrkin_Maier_2011}.
After some investigation of the convergence property and the size of systematic
errors, we found that $\mu^* = 2.86/x$ for the scalar and $\mu^* = 5.47/x$ for the
vector are optimal choices.
As shown in the right panel of Fig.~\ref{fig:ss},
the convergence is much better and the difference among 2-, 3- and 4-loop
is hardly visible for $x < 0.5$ fm.

The lattice correlators of the scalar and pseudoscalar chnnels are also plotted
on the right panel of Fig.~\ref{fig:ss}.
These correlators are multiplied by
$Z_{\rm S}^{\rm\overline{MS}} (\rm2\ GeV)^2$, the renormalization factor
of the scalar density determined in Sec.~\ref{sec:renormalization}.
Lattice result and the continuum perturbation theory agree very well in
$x \lesssim 0.25$ fm.
Significant difference found in $x > 0.25$ fm is due to non-perturbative
effects as discussed in the following sections.

\section{Non-perturbative effect on current correlators}

\begin{figure}[tb]
\begin{center}
\includegraphics[width=10cm]{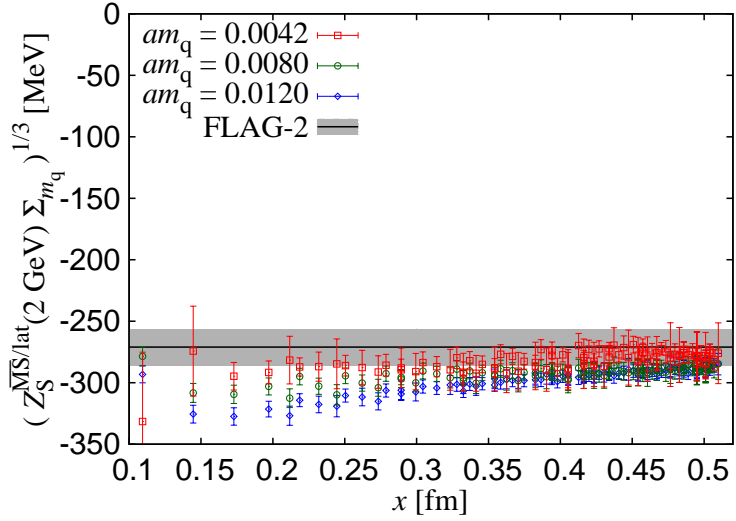}
\caption{
The cubic root of $\Sigma_{m_\q}$ renormalized by
$Z_{\rm S}^{\rm\overline{MS}}(2\rm\ GeV)$.
Input volume, $\beta$, and $m_s$ are same as Fig.~1 but result of three
valence masses are shown.
The solid line and the gray band stand for the FLAG average of the chiral
condensate at 2 GeV in the $\rm\overline{MS}$ shcme.
}
\label{fig:efcc}
\end{center}
\end{figure}

We discuss non-perturbative effects on the vector and axial-vector
correlators.
Using the PCAC relation, one can
relate V/A correlators to the chiral condensate \cite{Becchi:1980vz,Jamin:1992se}.
When the valence masses are degenerate, the relation is written as
\begin{equation}
\Sigma_{m_\q}(x)\equiv
-{\pi^2\big(Z_\V^{\rm\overline{MS}}\big)^2\over2m_\q}x^2
x_\nu\del_\mu\big(\Pi_{\A,\mu\nu}(x)-\Pi_{\V,\mu\nu}(x)\big)
=\langle\bar\q\q\rangle + O(m_\q).
\end{equation}
Here, $Z_{\rm V}^{\rm\overline{MS}}$ is to renormalize the vector and axial-vector
currents constructed by the local operators on the lattice.
The subtraction of $\del_\mu\Pi_{\V,\mu\nu}$, which vanishes in the continuum
theory, is to cancel the discretization effects.
Figure~\ref{fig:efcc} shows the lattice results of $\Sigma_{m_\q}(x)$ renormalized by
$Z_{\rm S}^{\rm\overline{MS}}(2\rm\ GeV)$ for three input masses,
corresponding to $M_\pi\sim300$, 400, and 500 MeV.
The lattice data would be flat in the chiral limit and coincide with the gray band,
which is the FLAG average \cite{Aoki:2013ldr} for $n_f = 2+1$,
$(\langle\bar\q\q\rangle^{\rm\overline{MS}}(2\rm\ GeV))^{1/3} = -271(15)$ MeV.

In contrast, the prediction for the scalar and pseudoscalar correlators from
OPE in literature fails to reproduce the lattice data.
It becomes prominent when one compares the difference of
these channels, {\it i.e.}
the lattice result shows larger difference than the OPE prediction.
Such inconsistency has been known
\cite{Novikov:1981xi,Shuryak:1993kg,Chu:1993cn,Faccioli:2003qz},
and a possible description by the instanton-induced 't Hooft interaction
is suggested.

\section{Renormalization of quark currents}
\label{sec:renormalization}

\begin{figure}[tb]
\begin{center}
\includegraphics[width=10cm]{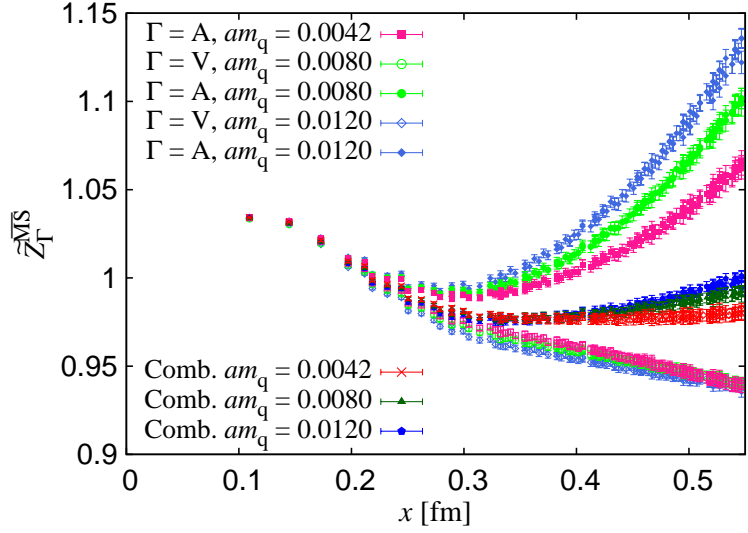}
\caption{
$\widetilde Z_{\rm V/A}^{\rm\overline{MS}}$ with the dependence on $x$,
calculated by (4.1) (square, circle, diamond)
and by taking the combination ${1\over8}(5\Pi_\V+3\Pi_\A)$
(cross, triangle, pentagon).
The results of same ensembles as Fig.~3 are shown.
}
\label{fig:Zv_mass}
\vspace{-3.5mm}
\end{center}
\end{figure}

In this section, we report the renormalization factors of quark currents
using correlators.
On each ensemble, the renormalization condition is applied by
\begin{equation}
\Big(\widetilde Z_\Gamma^{\overline{\rm MS}}\Big)^2
\Pi_\Gamma^{\rm lat}(x) = \Pi_\Gamma^{\rm\overline{MS}}(x),
\label{eq:renorm_cond}
\end{equation}
at a certain distance $x$.
The data of $\widetilde Z_\Gamma^{\rm\overline{MS}}$ may show some
$x$-dependence reflecting the discretization effects and non-perturbative
effects contained in $\Pi_\Gamma^{\rm lat}$.
As Fig.~\ref{fig:Zv_mass} shows, both the $x$-dependence and
mass-dependence of $\widetilde Z_{\rm V/A}^{\rm\overline{MS}}$
are found to be significant.
We take account of the non-perturbative effects
using the OPE.
The vector and axial-vector correlators are expressed
as a linear combination of vacuum expectation values of local operators,
\begin{equation}
\Pi_{\rm V/A}(x)
= {c_0\over x^6}
+ {c_{4,\bar\q\q}^{\V/\A}m_\q\langle\bar\q\q\rangle
+ c_{4,\rm G}^{\V/\A}\langle{\rm GG}\rangle\over x^2} + \cdots.
\end{equation}
Because of the relation $c_{4,\bar\q\q}^\V/c_{4,\bar\q\q}^\A=-3/5$
\cite{Shifman:1978bx,Jamin:1992se},
the combination ${1\over8}(5\Pi_\V+3\Pi_\A)$
cancels the bulk of the contribution from the chiral condensate
$\langle\bar\q\q\rangle$.
This combination $\widetilde Z_{\rm 5V+3A}^{\rm\overline{MS}}$ is
shown by crosses, triangles, and pentagons for different quark masses
in Fig.~\ref{fig:Zv_mass},
where the dependences on $x$ and on valence masses are dramatically reduced.
Although there is no mass dependence remains for the operators
of dimension four, the mass dependence of the data is still seen at $x > 0.4$ fm,
which may be attributed to the contributions from higher dimensional operators,
including the four-quark condensate $\langle\bar\q\bar\q\q\q\rangle$ and
$m_\q^2\langle\rm GG\rangle$.

\begin{figure}[t]
\begin{center}
\vspace{-1mm}
\includegraphics[width=10cm]{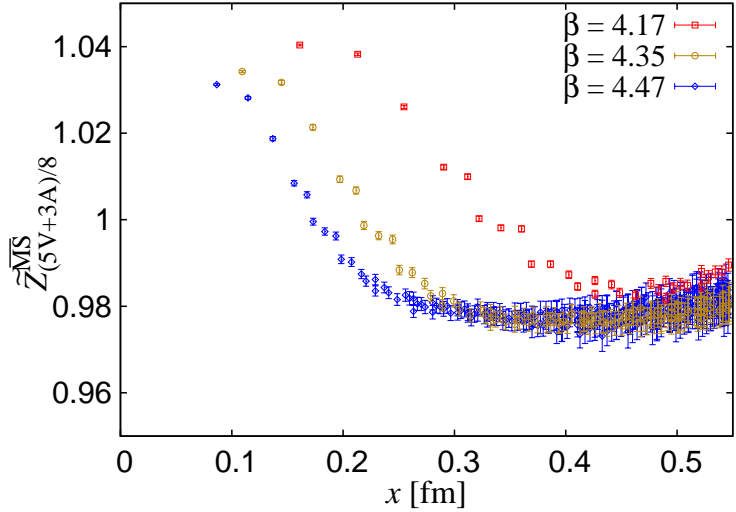}
\caption{
$\widetilde Z_{(\rm 5V+3A)/8}^{\rm\overline{MS}}(x)$ for each $\beta$
at similar pion masses $M_\pi \simeq 300$ MeV.
}
\label{fig:Zv_beta}
\vspace{-3.5mm}
\end{center}
\end{figure}

Figure~\ref{fig:Zv_beta} shows $\widetilde Z_{(\rm 5V+3A)/8}^{\rm\overline{MS}}(x)$
at each $\beta$.
The position where $\widetilde Z_{\rm (5V+3A)}^{\rm\overline{MS}}$ starts
deviating from a constant toward short-distance moves as the lattice spacing,
indicating the discretization effect.
The most significant discretization effect is of $O(a^2)$ which appears as $(a/x)^2$.
Since we have already subtracted the discretization effects at the tree level,
the remaining effect is $\alpha_s(a/x)^2$.

Taking account of these non-perturbative effects and discretization effects,
we determine the renormalization factor $Z_\V^{\rm\overline{MS}}$ in the
massless limit by a simultaneous fit of all ensembles using the fit function
\begin{equation}
\widetilde Z_{(5\V+3\A)/8}^{\rm\overline{MS}}(a;x)
= Z_\V^{\rm\overline{MS}}(\beta) + C_{-2} \alpha_s(a) (a/x)^2 + C_{4,\rm G}x^4
+ (C_{6,\q} + C_{6,\rm G} m_\q^2)x^6.
\label{eq:Zv}
\end{equation}
Our results are $Z_\V^{\rm\overline{MS}} = 0.951(4), 0.956(3), 0.961(3)$
at $\beta = 4.17, 4.35, 4.47$, respectively.

Determination of $Z_{\rm S}$ is more complicated
due to the possible instanton-induced effect.
Since instantons affect the scalar and pseudoscalar correlators to the opposite
direction with the same magnitude,
the na\"\i ve average ${1\over2}(\Pi_{\rm S}+\Pi_{\rm P})$ is independent of
such an effect.
According to the OPE, the average contains the contribution of the chiral
condensate, which may be cancelled by the difference between the vector and
axial-vector correlators.
Taking a combination of $\Pi_{\rm S}+\Pi_{\rm P}$ and $\Pi_\V-\Pi_\A$
to cancel these known $x$-dependence,
we are able to fit the Z-factor using the fit form similar to (\ref{eq:Zv}).
We obtain $Z_{\rm S}^{\rm\overline{MS}}(2\rm\ GeV) = 1.024(15), 0.922(11), 0.880(7)$
at $\beta = 4.17, 4.35, 4.47,$ respectively.

\section{Summary}

The main purpose of this work is to understand how precisely the continuum
theory predicts the short-distance behavior of current correlators
by comparing lattice results with the continuum theories.
The vector and axial-vector correlators agree with
the OPE in the region $x \lesssim 0.5{\rm\ fm}$,
while the scalar and pseudoscalar channels need more study to understand.

Controlling non-perturbative effects and discretization effects,
we determine renormalization factors of quark currents using correlators.
This procedure enables us to renormalize in a gauge invariant manner and to perform
the perturbative matching to the four-loop level.

\begin{acknowledgments}
Numerical simulations are performed on the IBM System Blue Gene Solution at High Energy
Accelerator Research Organization (KEK) under a support of
its Large Scale Simulation Program
(No. 13/14-04, 14/15-10).
We thank P. Boyle for the optimized code for BGQ.
This work is supported in part by the Grant-in-Aid of the Japanese Ministry of
Education (No. 25800147, 26247043, 15K05065) and the SPIRE (Strategic Program
for Innovative Research) Field5 project.
\end{acknowledgments}


\begin{thebibliography}{99}
\bibitem{Chetyrkin_Maier_2011}
K.G. Chetyrkin and A. Maier,
Nucl. Phys. B {\bf 844} (2011) 266
[arXiv:1010.1145 [hep-ph]].

\bibitem{Shifman:1978bx}
M.~A.~Shifman, A.~I.~Vainshtein and V.~I.~Zakharov,
Nucl.\ Phys.\ B {\bf 147} (1979) 385.

\bibitem{Martinelli_etal:1997}
G. Martinelli, G.C. Rossi, C.T. Sachrajda, S.R. Sharpe, M. Talevi and M. Testa,
Phys. Lett. B {\bf 411} (1997) 141
[hep-lat/9705018].

\bibitem{V.Gimenez_etal:2004}
V. Gim\'{e}nez, L. Giusti, S. Guerriero, V. Lubicz, G. Martinelli, S. Petrarca,
J. Reyes, B. Taglienti and E. Trevigne,
Phys. Lett. B {\bf 598} (2004) 227
[hep-lat/0406019].

\bibitem{K.Cichy_etal:2012}
K. Cichy, K. Jansen and P. Korcyl,
Nucl. Phys. B {\bf 865} (2012) 268
[arXiv:1207.0628 [hep-lat]].

\bibitem{Martinelli:1994ty}
G.~Martinelli, C.~Pittori, C.~T.~Sachrajda, M.~Testa and A.~Vladikas,
Nucl.\ Phys.\ B {\bf 445} (1995) 81
[hep-lat/9411010].

\bibitem{Fahy_etal:2015}
B. Fahy, G. Cossu, S. Hashimoto, T. Kaneko, J. Noaki and M. Tomii,
PoS LATTICE 2015,074

\bibitem{Chetyrkin_etal_1997}
K.G. Chetyrkin, B.A. Kniehl and M. Steinhauser,
Phys. Rev. Lett. {\bf 79} (1997) 2184
[hep-ph/9706430].

\bibitem{K.G.Chetyrkin_1997}
K.G. Chetyrkin,
Phys. Lett. B {\bf 404} (1997) 161
[hep-ph/9703278].

\bibitem{J.A.M.Vermaseren_etal:1997}
J.A.M. Vermaseren, S.A. Larin and T. van Ritbergen,
Phys. Lett. B {\bf 405} (1997) 327
[hep-ph/9703284].

\bibitem{Becchi:1980vz}
C.~Becchi, S.~Narison, E.~de Rafael and F.~J.~Yndurain,
Z.\ Phys.\ C {\bf 8} (1981) 335.

\bibitem{Jamin:1992se}
M.~Jamin and M.~Munz,
Z.\ Phys.\ C {\bf 60} (1993) 569
[hep-ph/9208201].

\bibitem{Aoki:2013ldr}
S.~Aoki {\it et al.},
Eur.\ Phys.\ J.\ C {\bf 74} (2014) 2890,
[arXiv:1310.8555 [hep-lat]].

\bibitem{Novikov:1981xi}
V.A.~Novikov, M.A.~Shifman, A.I.~Vainshtein and V.I.~Zakharov,
Nucl.\ Phys.\ B {\bf 191} (1981) 301.

\bibitem{Shuryak:1993kg}
E.~V.~Shuryak,
Rev.\ Mod.\ Phys.\  {\bf 65} (1993) 1.

\bibitem{Chu:1993cn}
M.C.~Chu, J.M.~Grandy, S.~Huang and J.W.~Negele,
Phys.\ Rev.\ D {\bf 48} (1993) 3340
[hep-lat/9306002].

\bibitem{Faccioli:2003qz}
P.~Faccioli and T.A.~DeGrand,
Phys.\ Rev.\ Lett.\  {\bf 91} (2003) 182001
[hep-ph/0304219].

\end{thebibliography}
\end{document}